\documentclass[journal=jpclcd, manuscript=letter]{achemso}

\usepackage[version=3]{mhchem} % Formula subscripts using \ce{}

\usepackage{graphicx}% Include figure files
\graphicspath{{./figures/}}
\usepackage{dcolumn}% Align table columns on decimal point
\usepackage{bm}% bold math
\usepackage{xspace}
\usepackage{siunitx}
\usepackage{hyperref}
\usepackage{braket}
\usepackage[nolist]{acronym}
\usepackage{subcaption}
\usepackage[inline]{enumitem}
\usepackage{simplewick}
\usepackage{multirow}
\usepackage{xcolor}
\usepackage{verbatim}
\usepackage{booktabs}
\usepackage{siunitx}

\begin{acronym}
  \acro{LGPL}{GNU Lesser General Public Licence}
  \acro{WF}{wave function}
  \acro{WFT}{wave function theory}
  \acro{LHS}{left-hand side}
  \acro{CC}{coupled cluster}
  \acro{CCS}{coupled cluster with single substitutions}
  \acro{CCSD}{coupled cluster with single and double substitutions}
  \acro{CCSD(T)}{CCSD with perturbative triples correction}
  \acro{CCSDT}{coupled cluster with single, double and triple substitutions}
  \acro{CC2}{approximate coupled cluster singles and doubles}
  \acro{CC3}{approximate coupled cluster singles, doubles and triples}
  \acro{PT}{perturbation theory}
  \acro{MBPT}{many-body perturbation theory}
  \acro{MO}{molecular orbital}
  \acro{AO}{atomic orbital}
  \acro{MP}{M{\o}ller--Plesset}
  \acro{HF}{Hartree--Fock}
  \acro{QM}{quantum mechanics}
  \acro{QC}{quantum chemistry}
  \acro{RHS}{right-hand side}
  \acro{SCF}{self-consistent field}
  \acro{BCH}{Baker--Campbell--Hausdorff}
  \acro{MC}{Monte Carlo}
  \acro{FCI}{full configuration interaction}
  \acro{DMC}{diffusion Monte Carlo}
  \acro{QMC}{quantum Monte Carlo}
  \acro{VMC}{variational Monte Carlo}
  \acro{RI}{resolution-of-the-identity}
  \acro{FCIQMC}{full configuration interaction quantum Monte Carlo}
  \acro{SCCT}{stochastic coupled cluster theory}
  \acro{RDM}{reduced density matrix}
  \acro{MSQMC}{Model Space Quantum Monte Carlo}
  \acro{CCMC}{Coupled Cluster Monte Carlo}
  \acro{diagCCMC}{diagrammatic Coupled Cluster Monte Carlo}
  \acro{RHF}{Restricted Hartree--Fock}
  \acro{UHF}{Unrestricted Hartree--Fock}
\end{acronym}

\newcommand{\vect}[1]{\bm{#1}} % Vector
\newcommand{\hairsp}{\hspace{1pt}}% hair space
\newcommand{\ie}{\textit{i.\hairsp{}e.}\xspace}

\newcommand{\Det}[1]{D_{\textbf{#1}}}
\newcommand{\refd}[0]{\ket{\Det{0}}}
\newcommand{\sdet}[1]{\ket{\Det{#1}}}
\newcommand{\ccwav}[0]{\ket{\mathrm{CC}}}

\newcommand{\expT}[1]{\mathrm{e}^{#1}}
\newcommand{\tamp}[2]{t_{#1}^{#2}}
\newcommand{\tampc}[1]{t_{\textbf{#1}}}
\newcommand{\cluster}[2]{\tau_{#1}^{#2}}
\newcommand{\excitor}[1]{\tau_{\textbf{#1}}}
\newcommand{\oneBody}[2]{{h}_{#1}^{#2}}
\newcommand{\fock}[2]{{f}_{#1}^{#2}}
\newcommand{\twoBody}[4]{{g}_{#1 #2}^{#3 #4}}
\newcommand{\twoBodya}[4]{\bar{g}_{#1 #2}^{#3 #4}}
\newcommand{\prob}[1]{p_{\mathrm{#1}}}

\author{Charles J. C. Scott}
 \email{cjcargillscott@gmail.com}
 \affiliation{Department of Chemistry, University of Cambridge, Cambridge, UK}

\author{Roberto Di Remigio}
\email{roberto.d.remigio@uit.no}
 \affiliation{Hylleraas Centre for Quantum Molecular Sciences, Department of Chemistry, University of Troms{\o} - The Arctic University of Norway, N-9037 Troms{\o}, Norway}
 \alsoaffiliation{Department of Chemistry, Virginia Tech, Blacksburg, Virginia 24061, United States}

\author{T. Daniel Crawford}
 \affiliation{Department of Chemistry, Virginia Tech, Blacksburg, Virginia 24061, United States}

\author{Alex J. W. Thom}
 \affiliation{Department of Chemistry, University of Cambridge, Cambridge, UK}

\title{Diagrammatic Coupled Cluster Monte Carlo}

\begin{document}

\begin{tocentry}
  \centering
\includegraphics[width=5cm]{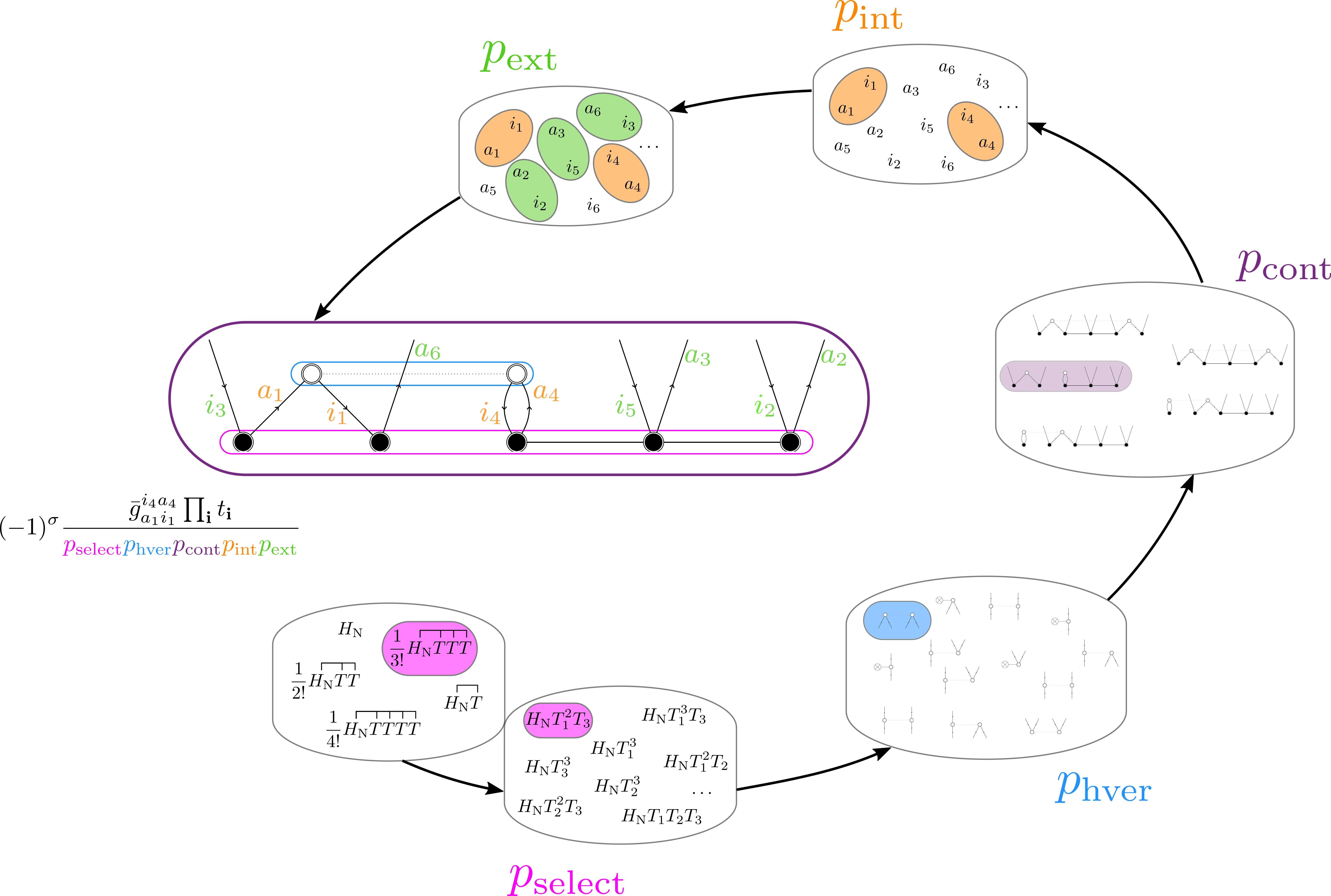}
\end{tocentry}

\begin{abstract}
We propose a modified coupled cluster Monte Carlo algorithm that stochastically
samples connected terms within the truncated Baker--Campbell--Hausdorff
expansion of the similarity transformed Hamiltonian by construction of coupled
cluster diagrams on the fly.
Our new approach -- diagCCMC -- allows propagation to be performed using only
the connected components of the similarity-transformed Hamiltonian, greatly
reducing the memory cost associated with the stochastic solution of the coupled
cluster equations.
We show that for perfectly local, noninteracting systems, diagCCMC is able to
represent the coupled cluster wavefunction with a memory cost that scales
linearly with system size. The favorable memory cost is observed with the only
assumption of fixed stochastic granularity and is valid for arbitrary levels of
coupled cluster theory. Significant reduction in memory cost is also shown to
smoothly appear with dissociation of a finite chain of helium atoms.
This approach is also shown not to break down in the presence of strong
correlation through the example of a stretched nitrogen molecule.
Our novel methodology moves the theoretical basis of coupled cluster Monte Carlo
closer to deterministic approaches.
\end{abstract}

Over the last half-century the \ac{CC} wavefunction Ansatz has proved remarkably effective at
representing the solution of the Schr\"{o}dinger equation in a polynomial
scaling number of parameters while providing size-extensive and -consistent
results.
Despite reducing the \ac{FCI} $N!$ factorial scaling to
polynomial, the computational cost of \ac{CC} methods, measured in terms of both
required CPU floating-point operations and memory, is still an issue.
The \ac{CCSD} and \ac{CCSD(T)} approximations provide a balance between
computational cost and accuracy that has led to relatively wide adoption,
but are eventually precluded for many large systems.

Recent work has made great progress on this issue through application of various
approximations, which enable calculations to be performed with reduced memory
and computational costs. In particular, various approximations exploiting the
locality of electron correlation allow calculations
with costs asymptotically proportional to measures of system
size. These include approaches based on orbital localisation\cite{Pulay1983-ht,
	Stoll92, Saebo1993-qx, Hampel1996-yy, Schutz99, Schutz00:Ta, Schutz00:Tb,
	Schutz2001-da, Schutz02, Mata07, Taube08, Piecuch09, Neese09, Neese2009-ch,
	Piecuch10, Piecuch10b, Liakos11, Tew11, Yang11, Yang12, Liakos12, Hattig12,
	Krause12, Masur13, Riplinger2013-mz, Riplinger2013-ue,
	Werner15, Liakos2015-yw, Riplinger2016-zo, Pavosevic2016-kw, Pavosevic2017-tl,
	Saitow2017-rr, Guo2018-bg, Yang2012-je, Schwilk2015-iv, Ma2017-rl,
	Schwilk2017-zf, Pavosevic17, Ma2018-jb}, molecular fragmentation\cite{Kitaura99,
	Li04, Fedorov05, Li06, Li09, Stoll09, Friedrich09, Ziolkowski2010-oo,
	Kristensen2011-ck, Rolik2011-fn, Hoyvik2012-td, Gordon12, Eriksen2015-kw}, and
decompositions, such as \acl{RI}, Cholesky or singular-value, of the two-electron
integrals tensors\cite{Kinoshita03, Yang11, Yang12, Hattig00, Hattig02, Koch03,
	Pedersen04, Epifanovsky2013}. However, while providing large efficiencies in
\ac{CCSD} calculations, higher truncation levels will generally exceed available
memory resources before such approximations are a reasonable proposition.

In this letter we propose and demonstrate a \acl{CC}-based projector \ac{MC}
algorithm that enables automatic exploitation of the wavefunction sparsity
for arbitrary excitation orders.
Our methodology can be particularly beneficial for
localised representations of the wavefunction, but it is not limited by
assumptions of locality.
The approach can fully leverage the sparsity inherent in the \ac{CC} amplitudes
at higher excitation levels,\cite{Lehtola2017} allowing dramatic reductions in
memory costs for higher levels of theory.

The \ac{CC} wavefunction is expressed as an exponential
transformation of a reference single-determinant wavefunction $\refd$:
\begin{equation}\label{eq:standard-cc-wav}
\ccwav = \expT{T} \refd
\end{equation}
where the cluster operator $T$ is given as a sum of second-quantised excitation operators:
\begin{equation}\label{eq:cluster-op}
T = \sum_{k} T_{k},
\end{equation}
with the $k$-th order cluster operators expressed as sums of excitation operators weighted
by the corresponding \emph{cluster amplitudes}:
\begin{equation}\label{eq:cluster-k}
T_{k} = \sum_{\textbf{i}\in k^{\text{th}}\,\text{replacements}}\tampc{i}\excitor{i}
= \frac{1}{(k!)^2} \sum_{\substack{a_1, a_2, \ldots, a_k \\ i_1, i_2, \ldots, i_k}}
\tamp{a_1a_2\ldots a_k}{i_1i_2\ldots i_k}
\cluster{i_1i_2\ldots i_k}{a_1a_2\ldots a_k},
\end{equation}
in the tensor notation for second quantisation proposed by \citet{Kutzelnigg1997-vt}.
%An overview of our nomenclature and notation is summarised in Table \ref{tab:nomenclature}.
Upon truncation of the cluster operator to a certain excitation level $l$ and
projection of the Schr\"{o}dinger equation onto the corresponding excitation
manifold one obtains the \emph{linked} energy and cluster amplitudes equations:
\begin{subequations}\label{eq:linked-cc}
	\begin{align}
	\braket{\Det{0} | \bar{H}_\mathrm{N} | \Det{0}} &= E_\mathrm{CC} \\
	\Omega_{\textbf{n}}(\vect{t}) = \braket{\Det{n} | \bar{H}_\mathrm{N} | \Det{0}} &= 0.
	\end{align}
\end{subequations}
We have introduced the similarity-transformed Hamiltonian,
$\bar{H}_\mathrm{N}=\expT{-T}H_\mathrm{N}\expT{T}$, and $\sdet{n}$ can be any
state within the projection manifold (up to an $l$-fold excitation of $\refd$).
These \ac{CC} equations are manifestly size-extensive order-by-order and
term-by-term and furthermore provide the basis for the formulation of response
theory.\cite{Christiansen1998-pe}

\ac{CC} methods have to be carefully derived order-by-order and
their implementation subsequently carried out, a process that can be rather
time-consuming and error-prone.\cite{Noga1987-an, Kucharski1992-pe, Hirata2000-sx}
It has long been recognised that the use of normal-ordering, \cite{Crawford2000, Shavitt2009-mr}
Wick's theorem,\cite{Wick1950-iy} and the ensuing diagrammatic
techniques\cite{Kucharski1986-sz} can be leveraged to automate both
steps\cite{Harris1999-gf, Crawford2000, Kallay2001-yv, Kallay2004-ug,
	Kallay2003-qi, Kallay2004-fv, Lyakh2005-do, Krupicka2017-rb}, though
spin-adaptation can still pose significant challenges.\cite{Matthews2013-cz, Matthews2015-zd, Wang2018-oh}
Consider the normal-ordered, electronic Hamiltonian:
\begin{equation}\label{eq:electronic-hamiltonian-N}
\begin{aligned}
H_\mathrm{N} &= F + \Phi = \sum_{pq} \fock{p}{q}e_{q}^{p} + \frac{1}{2}\sum_{pqrs}\twoBody{p}{q}{r}{s}e_{rs}^{pq} \\
&= \sum_{pq} \fock{p}{q}e_{q}^{p} + \frac{1}{4}\sum_{pqrs}\twoBodya{p}{q}{r}{s}e_{rs}^{pq} = H - E_{\mathrm{ref}},
\end{aligned}
\end{equation}
its similarity transformation admits a \ac{BCH} expansion truncating exactly
after the four-fold nested commutator.\cite{Helgaker2000-yb, Shavitt2009-mr}
%\begin{equation}\label{eq:H-commutator-expansion}
%\begin{aligned}
%  \bar{H}_\mathrm{N} &= \sum^4_{k = 0} \frac{1}{k!} \nestcommR{H_\mathrm{N}}{T}{k} \\
%  &= H_\mathrm{N} + \BCHfirst{H_\mathrm{N}}{T} + \BCHsecond{H_\mathrm{N}}{T} \\
%&+ \BCHthird{H_\mathrm{N}}{T} + \BCHfourth{H_\mathrm{N}}{T},
%\end{aligned}
%\end{equation}
Since all excitation operators are normal-ordered and commuting, the commutator expansion
lets us reduce the Hamiltonian-excitation operator products to only those terms which are
\emph{connected}.\cite{Crawford2000, Shavitt2009-mr}
Excitation operators will only appear to the \emph{right} of the Hamiltonian and only terms
where each excitation operator shares at least one index with the Hamiltonian will lead to
nonzero terms in the residuals $\Omega_{\textbf{n}}(\vect{t})$ appearing in
eqs.~(\ref{eq:linked-cc}):
\begin{equation}\label{eq:H-connected-expansion}
\begin{aligned}
\bar{H}_\mathrm{N} &= (H_\mathrm{N}\expT{T})_\mathrm{c} = H_\mathrm{N}
% First-order
+ \contraction{}{H_\mathrm{N}}{}{T}
H_\mathrm{N}T
% Second-order
+ \frac{1}{2!}
\contraction{}{H_\mathrm{N}}{}{T}
\contraction{}{H_\mathrm{N}}{T}{T}
H_\mathrm{N}TT \\
% Third-order
&+ \frac{1}{3!}
\contraction{}{H_\mathrm{N}}{}{T}
\contraction{}{H_\mathrm{N}}{T}{T}
\contraction{}{H_\mathrm{N}}{TT}{T}
H_\mathrm{N}TTT
% Fourth-order
+ \frac{1}{4!}
\contraction{}{H_\mathrm{N}}{}{T}
\contraction{}{H_\mathrm{N}}{T}{T}
\contraction{}{H_\mathrm{N}}{TT}{T}
\contraction{}{H_\mathrm{N}}{TTT}{T}
H_\mathrm{N}TTTT.
\end{aligned}
\end{equation}
Moreover, by virtue of Wick's theorem\cite{Wick1950-iy, Kutzelnigg1997-vt}, the
products of normal-ordered strings appearing in the connected expansion will
still be expressed as normal-ordered strings, further simplifying the algebra.
The requirement of shared indices between the Hamiltonian and cluster
coefficients enables the resulting equations to be solved via a series of tensor
contractions between multi-index quantities: the sought-after cluster amplitudes
and the molecular one- and two-electron integrals.
The iterative process required to solve eqs.~(\ref{eq:linked-cc}) is highly
amenable for a rapid evaluation on conventional computing
architectures,\cite{Stanton1991-dc, Gauss1991-wo}
but remains non-trivial to parallelise,\cite{Peng2016-oy} especially for higher truncation
orders in the \ac{CC} hierarchy.\cite{Matthews2015-zd}
A proper factorisation of intermediates is essential to achieve acceptable time
to solution and memory requirements.

In recent years some of us have been involved in developing
a projector \ac{MC} algorithm to obtain the \ac{CC} solutions within a
stochastic error bar.\cite{Thom2010, Spencer2015a, Franklin2016, Scott2017}
The starting point, as with any projector \ac{MC} method, is the imaginary-time
Schr\"{o}dinger equation\cite{Booth2009, Foulkes2001-lh, Toulouse2015-yz}
obtained after a Wick rotation $\tau \leftarrow \mathrm{i} t$.
Repeated application of the approximate linear propagator to a trial
wavefunction will yield the ground-state solution:
\begin{equation}
\Ket{\Psi(\tau+\delta\tau)} = \left[ 1 - \delta\tau (H - S) \right]\Ket{\Psi(\tau)}
\end{equation}
where $S$ is a free parameter that is varied to keep the normalisation of $\Psi(\tau)$
approximately constant.
In the \ac{CC}\ac{MC} and \ac{FCIQMC} approaches, a population of particles in
Fock space represents the wavefunction and evolves according to simple rules of
spawning, death, and annihilation.\cite{Booth2009, Thom2010}
For a \ac{CC} Ansatz, unit particles may represent nonunit contributions to
\ac{CC} amplitudes by letting the intermediate normalisation condition vary with
the population on the reference determinant: $\braket{\Det{0} |
	\mathrm{CCMC}(\tau)} = N_0(\tau)$. A factor of $\frac{1}{N_0(\tau)}$ is
removed from the definition of $T(\tau)$ and this determines the \emph{granularity} of
amplitude representation: amplitude values smaller than $\frac{1}{N_0(\tau)}$
are stochastically rounded during the calculation, \emph{vide infra}.
To avoid confusion, we denote the so-modified cluster operators and amplitudes
as $T^\prime$ and $\tampc{n}^\prime$, respectively, so $\ket{\mathrm{CCMC}} = N_0\expT{\frac{T^\prime}{N_0}}\refd$.
Thus, in the \emph{unlinked} formulation first put forward by Thom,\citep{Thom2010}
the dynamic equation for the amplitudes becomes:
\begin{equation}\label{eq:unlinked-update}
\tampc{n}^\prime \rightarrow \tampc{n}^\prime- \delta \tau
\braket{\Det{0} | \excitor{n}^\dagger [H-S] | \mathrm{CCMC}},
\end{equation}
where we have dropped the $\tau$-dependence for clarity.
\ac{CC}\ac{MC} is fully general with respect to the truncation
level in the cluster operator and sidesteps the need to store a full
representation of the wavefunction at any point.
\ac{CC}\ac{MC} should allow for the effective solution of the \ac{CC} equations
with a much reduced memory cost, as previously realised in the \ac{FCIQMC}
method.\cite{Booth2009, Cleland2010, Petruzielo2012, Blunt2015-iq}
However, while various cases demonstrate memory cost reduction, especially in
the presence of weak correlation,\cite{Neufeld2017} the corresponding increase
in computational cost was large even by the standards of projector \ac{MC}
methods and modifications used in related approaches, such as the initiator
approximation,\cite{Cleland2010} proved comparatively
ineffective.\cite{Spencer2015a}

Combination with the linked \ac{CC} formulation seems to be one possible
remedy for these issues and is furthermore the basis for decades of theoretical and
implementation work in the \emph{deterministic} community. \citet{Franklin2016}
have discussed a \ac{CC}\ac{MC} algorithm to sample eqs.~(\ref{eq:linked-cc})
using the update step:
\begin{subequations}
	\begin{align}
	\tampc{n}^\prime &\rightarrow \tampc{n}^\prime - \delta \tau N_{{0}}\braket{\Det{0} | \excitor{n}^\dagger \bar{H} | \Det{0}},
	\,\, (\sdet{n} \neq \refd)  \label{eq:orig-linked-grad-update} \\
	N_{{0}} &\rightarrow N_{{0}} - \delta \tau N_{{0}} \braket{\Det{0} | \bar{H} - S | \Det{0}}
	\end{align}
\end{subequations}
The authors however noted that the use of the similarity-transformed Hamiltonian
required an \emph{ad hoc} modification:
\begin{equation}\label{eq:modified-linked-grad-update}
\tampc{n}^\prime \rightarrow \tampc{n}^\prime
- \delta \tau N_{{0}}\braket{\Det{0} | \excitor{n}^\dagger [\bar{H} - E_\mathrm{CC}]| \Det{0}}
- \delta\tau(E_{\mathrm{CC}} - S) \tampc{n},
\end{equation}
to deal with convergence issues with the projected energy prior to the
initialisation of population control.
In addition, due to evaluation of $\bar{H}$ via the commutator expansion of the
bare Hamiltonian, rather than the sum of
connected Hamiltonian-excitation operator products~(\ref{eq:H-connected-expansion}), some
disconnected terms were included.
These extraneous terms in the algorithm of
\citet{Franklin2016} have been observed to correctly cancel out on average, but
render unnecessarily complex the sampling of connected contributions only.
Eventually, it is difficult to develop stochastic counterparts to approximations,
such as the CCn hierarchy,\cite{Christiansen1995-vy, Koch1997-nm} proposed
within deterministic \ac{CC} theory.

We here reconsider the implementation of the linked \ac{CC}\ac{MC} algorithm in the
light of the diagrammatic techniques used in deterministic \ac{CC}, an approach
we name \ac{diagCCMC}.
The update equation can be easily derived as a finite difference approximation to the
exact imaginary-time dynamics of the coupled cluster wavefunction under the assumption
of \emph{constant} intermediate normalisation:
\begin{equation}\label{eq:linked-grad-update}
\tampc{n}(\tau + \delta\tau) = \tampc{n}(\tau) - \delta\tau \braket{\Det{0} | \excitor{n}^{\dagger} \bar{H}_{\mathrm{N}}(\tau) | \Det{0} }.
\end{equation}
This has been noted elsewhere,\cite{Ten-No2017} and we will discuss its
implications in greater detail in a subsequent communication\cite{diagccmc},
but for now it will suffice to observe that since this is a projector \ac{MC} approach
it will eventually converge to the lowest energy solution of the \ac{CC} equations.
The existence of multiple solutions to the nonlinear \ac{CC} equations is
well-documented,\cite{Kowalski1998a,Kowalski1998b,Piecuch2000} and a projector
\ac{MC} approach could result in a different solution to the \ac{CC} equations
than the one found \emph{via} a deterministic procedure, where iteration
stabilises upon whichever solution is approached first from a given starting
point.
In practice a difference is only observed if a highly truncated form of \ac{CC} has
been applied inappropriately to a system, and even then only in the worst cases.

The second term on the right-hand side is the contribution to the \ac{CC} vector
function $\Omega_{\textbf{n}}(\mathbf{t})$ resulting from the projection upon the determinant
$\sdet{n}$ and is representable as a finite sum of enumerable diagrams.
Thus, at each iteration, we wish to randomly select $n_{\mathrm{a}}$ diagrams
from $\braket{\Det{0} | \excitor{n}^\dagger \bar{H}_{\mathrm{N}} | \Det{0}}$.
Each of these will be in the form of an excitation operator, $\excitor{\textbf{i}}$, and corresponding
weight, $w_{\textbf{i}}$, selected with some known, normalised probability,
$\prob{diagram}$, such that we expect to select any given
contributing diagram $\prob{diagram}\times n_a$ times at each iteration.
As by construction $\braket{\Det{0} | \excitor{j}^{\dagger}\excitor{i}| \Det{0}} = \delta_{\textbf{ij}}$,
a selected term can be found to contribute to the update of a single coefficient
with no additional sign considerations.
Rather than explicitly introduce a particulate representation of the
coefficients, as in \ac{FCIQMC} and previous \ac{CC}\ac{MC} approaches, we
stochastically round all coefficients $t_{\textbf{n}}$ with magnitude below some
strictly positive \emph{granularity} parameter $\Delta$.
If $|t_{\textbf{n}}| < \Delta$, then $|t_{\textbf{n}}|$ is rediscretised to
either $\Delta$ (with probability $\left| \frac{t_{\textbf{n}}}{\Delta}
\right|$) or 0 (with probability $1-\left| \frac{t_{\textbf{n}}}{\Delta}
\right|$).\cite{Overy2014,diagccmc} This can be shown to be equivalent to a
representation with unit particles and constant intermediate normalisation
$\frac{1}{\Delta}$.

\begin{figure}[htb]
\centering
\includegraphics[width=.8\linewidth]{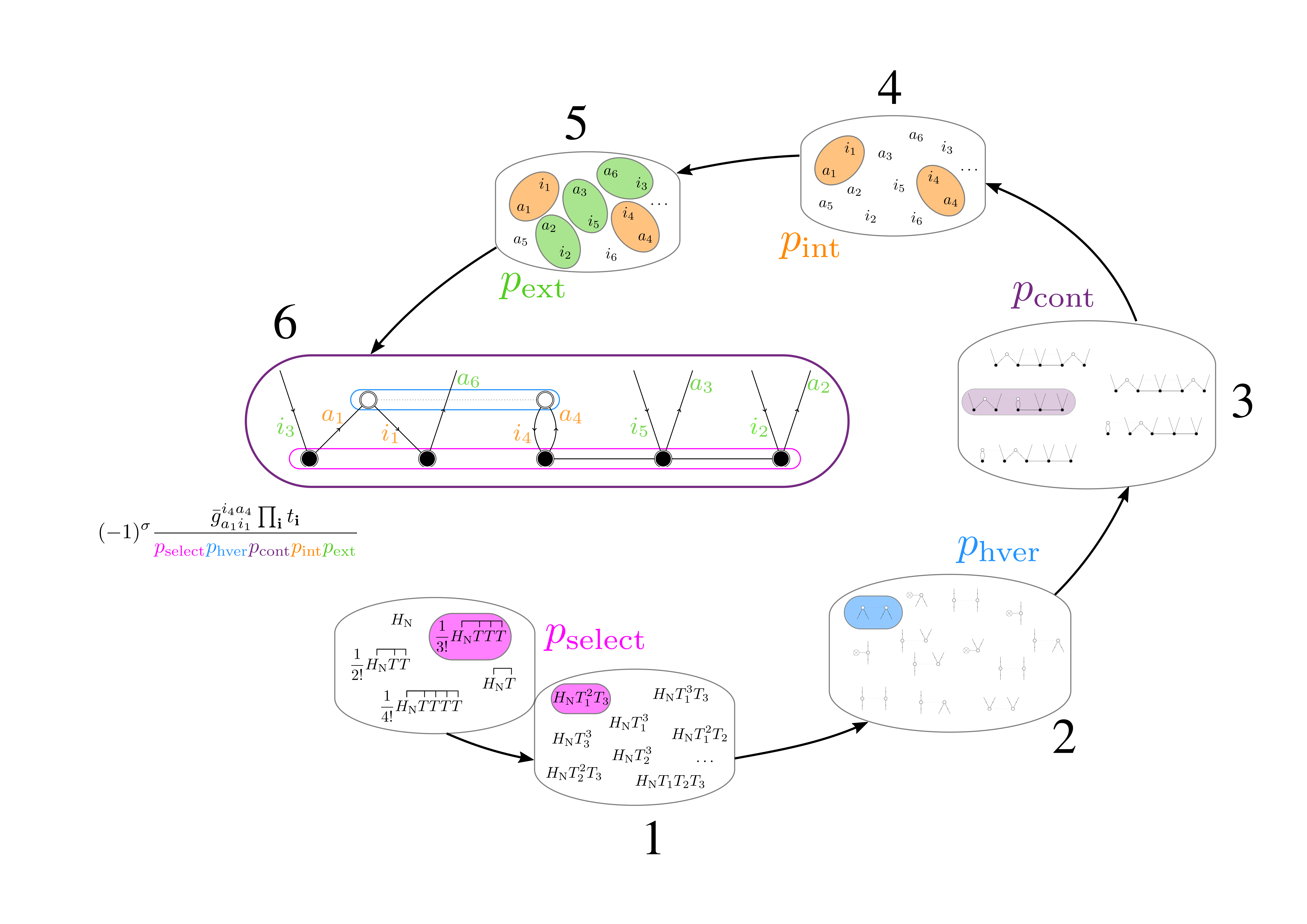}
\caption{Graphical depiction of the diagrammatic \ac{CC}\ac{MC} algorithm.
  This example shows the steps involved in the generation of one of the possible
  diagrams contributing to the $T_{3}$ equations.}
\label{fig:T12-T3-example}
\end{figure}

We perform diagram selection by reading off terms from right-to-left in
$\braket{\Det{0} | \excitor{n}^\dagger \bar{H}_{\mathrm{N}} | \Det{0}}$:
\begin{enumerate}
\item Select a random cluster of excitation operators with probability $\prob{select}$ utilising the even selection scheme\cite{Scott2017} restricted to clusters of
  at most 4 excitation operators. This corresponds to simultaneously selecting a term in the \ac{BCH}
  expansion~(\ref{eq:H-connected-expansion}) and the excitation level of each
  excitation operator in the commutator.
\item Select one of the 13 possible $H_N$ vertices\cite{Crawford2000, Shavitt2009-mr} with some probability $\prob{hvertex}$.
\item Select the contraction pattern of the chosen cluster and Hamiltonian
  vertex. This identifies a specific Kucharski--Bartlett sign
  sequence\cite{Kucharski1986-sz, Crawford2000, Shavitt2009-mr} for the diagram
  we are considering and which excitation operators are associated with which term within
  the sign sequence with probability $\prob{contract}$.
\item Select which indices of each excitation operator will be contracted with the
  Hamiltonian vertex. Having selected the contraction pattern this is a matter
  of simple combinatorics, with a given set of indices selected with probability
  $\prob{internal}$.
\item Select the external indices of the Hamiltonian vertex with probability $\prob{external}$.
\item Evaluate the index of resulting projection determinant in the update step,
\ie $\bra{\Det{0}}\excitor{n}^\dagger$, and the diagrammatic amplitude including
all parity factors.
\end{enumerate}
This obtains a single specific diagram with probability:
\begin{equation}
\prob{diagram} = \prob{select}\prob{hver}\prob{cont}\prob{int}\prob{ext},
\end{equation}
where the obvious abbreviations have been used to refer to each of the
previously stated probabilities. These are \emph{conditional}
probabilities, as the various events leading to the computed
$\prob{diagram}$ are not independent. This procedure to select diagrams
can be visualised as graphically building the diagram bottom-up,
see Figure~\ref{fig:T12-T3-example}.

To evaluate the contribution of a selected diagram to our propagation, we
slightly modify the standard rules of diagrammatic interpretation.
Instead of summing over all indices, and thus having to correct for any
potential double counting, our algorithm selects a \emph{specific} diagram along
with a \emph{specific} set of indices for all lines.

To ensure proper normalisation of our sampling probability, we require there be
only a single way to select diagrams related by:
\begin{itemize}
\item The antipermutation of antisymmetrised Goldstone vertex indices.
\item The antipermutation of cluster operator particle or hole indices.
\item The commutation of cluster operators.
\end{itemize}
All these modifications can be viewed as replacing sums
$\frac{1}{2}\sum\limits_{ij}$ with $\sum\limits_{i>j} + \frac{1}{2}\delta_{ij}$.
In the first two cases summation runs over equivalent indices and the $i=j$ term
must be zero, while in the third case summation runs over excitation operators and the $i=j$
term corresponds to a diagram with additional symmetry that as such must be
treated more carefully to ensure unique selection of a Kucharski--Bartlett sign
sequence\cite{Kucharski1986-sz, Crawford2000, Shavitt2009-mr}.
Specifically, we do not require an additional factor of $\frac{1}{2}$ for:
\begin{itemize}
\item Each pair of equivalent internal or external lines.
\item Two cluster operators of the same rank but with different specific indices, provided they have a well-determined ordering on selection.
\end{itemize}
Additionally, to include the effect of permutation operators $\hat{P}$ for
inequivalent external lines we must permute the hole and particle indices of a
resulting excitation operator to a unique antisymmetrised ordering for storage. This ensures
proper cancellation between all equivalent orderings, which could otherwise
differ due to the stochastic sampling.
Eventually, the amplitude of the contribution of the selected diagram,
$w_{\mathrm{diagram}}$, is given as the product of the cluster amplitude,
$w_{\mathrm{clus}} = \prod_{\textbf{i}}\tampc{i}$, and Hamiltonian element,
$w_{\mathrm{hamil}}$, with appropriately determined parity $(-1)^{\sigma}$.
The overall contribution of a single selected diagram to the coefficient
$\tampc{n}$ determined by the open lines of the diagram will be:
\begin{equation}\label{eq:diagrammatic-update-contribution}
\frac{w_{\mathrm{diagram}}}{\prob{diagram}} =\frac{(-1)^{\sigma}w_{\mathrm{clus}}w_{\mathrm{hamil}}}{\prob{select}\prob{hver}\prob{cont}\prob{int}\prob{ext}},
\end{equation}
wherever possible we aspire to have $\prob{diagram} \propto |w_{\mathrm{diagram}}|$.\cite{Scott2017}

%%RESULTS

We will now demonstrate the ability of diagCCMC to recover energies at high
levels of \ac{CC} theory on the nitrogen molecule in a stretched geometry
($r_{\mathrm{NN}} = 3.6\,a_{0}$).
It has previously been shown that connected contributions up to
hextuples are vital to obtaining high accuracy for this system.\cite{Chan2004}
Correlation energies for a range of basis sets and truncation levels are
reported in Table~\ref{tab:stretched-n2}, showing agreement within error bars
with deterministic results\cite{MRCC} and the existing literature in all but
the most extreme cases, where convergence to a different solution is observed
as noted previously.

We then turn our attention to test systems of beryllium and
neon atoms at a variety of truncation levels. Extending these systems by
introducing noninteracting replicas illustrates the behaviour of our approach in
the presence of locality in comparison to previous Fock-space stochastic
methods, namely the original unlinked \ac{CC}\ac{MC} (hereafter simply referred
to as \ac{CC}\ac{MC}) and \ac{FCIQMC}.

To allow reasonable comparison between \ac{diagCCMC}, \ac{CC}\ac{MC}, and \ac{FCIQMC}
all calculations were performed with:
\begin{itemize}
\item Granularity parameter $\Delta$ equal to $10^{-4}$. This is the threshold
  for the stochastic rounding of the cluster amplitudes.
\item $\delta\tau$ and $n_{\text{attempts}}$ such that, on each iteration, a
  spawning event may have maximum size of $3\times 10^{-4}$.
\end{itemize}
For \ac{CCMC} and \ac{FCIQMC} this corresponds to a stable calculation with
reference population of $N_0 = 10^{4}$ and a timestep such that no spawning
event produces more than three particles.
\ac{CC}\ac{MC} and \ac{FCIQMC} calculations were performed with the
HANDE-QMC code\cite{Spencer2015b,Spencer2019-tc} using the default, uniform
excitation generators.
For \ac{CC}\ac{MC}, we adopted the even selection scheme of \citet{Scott2017}.
The molecular integrals were generated in FCIDUMP format using the Q-Chem\cite{Shao2015-dy} and
Psi4\cite{Parrish2017-fd} quantum chemistry program packages, see the Supporting
Information for more details.\cite{Scott2019-ge}

\begin{table}[htb]
	\caption{Correlation energy for different levels of theory and basis sets
for \ce{N2} with $r_{\mathrm{NN}} = 3.6\,a_{0}$. Molecular integrals were
generated in FCIDUMP format with the Psi4 program package.\cite{Parrish2017-fd}
The STO-3G and 6-31G results were computed using MRCC\cite{MRCC}. The canonical
restricted \acl{HF} orbitals were used, giving $E_{\mathrm{ref}} =
\SI{-106.937562}{\hartree}$ and $\SI{-108.360046}{\hartree}$ in the STO-3G and
6-31G bases, respectively.}
	\label{tab:stretched-n2}
	\begin{tabular}{l r S S}
		\toprule
		                   &              & {STO-3G}    & {6-31G} \\
		\cline{2-4}
		\multirow{2}{*}{SD} & \ac{CC}       & -0.589163 & -0.491480 \\
		                   & \ac{diagCCMC} & -0.799$(2)^{a}$  & -0.4921(7) \\
		\cline{2-4}
		\multirow{2}{*}{SDT} & \ac{CC} & -0.589923 & -0.533600 \\
                           & \ac{diagCCMC} & -0.6092$(8)^{a}$  & -0.5341(9) \\
		\cline{2-4}
		\multirow{2}{*}{SDTQ} & \ac{CC}       & -0.523049 & \\
		                    & \ac{diagCCMC}  & -0.5244(9) & \hspace{1cm}\text{$\phantom{t}^{\text{b}}$} \\
		\cline{2-4}
		\multirow{2}{*}{SDTQ5} & \ac{CC} & -0.523036 &  \\
                             & \ac{diagCCMC}  & -0.5249(6) & \hspace{1cm}\text{$\phantom{t}^{\text{b}}$} \\
		\cline{2-4}
		\multirow{2}{*}{SDTQ56} & \ac{CC} & -0.527863 &  \\
                              & \ac{diagCCMC}  & -0.5271(8) & \hspace{1cm}\text{$\phantom{t}^{\text{b}}$} \\
		\bottomrule
	\end{tabular}

$\phantom{t}^{a}$In these cases the stochastic, imaginary-time propagation was
found to initially converge to the conventional \ac{CC} solution, before
relaxing to another, lower energy solution.\cite{Piecuch2000}\hfill

$\phantom{t}^{\text{b}}$ Value not computed due to computational constraints.\hfill
\end{table}

We report the correlation energies obtained for an isolated \ce{Be} atom and the
noninteracting replicas systems in Table~\ref{tab:ncombos-w-tl}. We compare \ac{CC} results
up to and including quadruple excitations with \ac{FCIQMC}.
For these systems CCSDTQ is equivalent to \ac{FCI}, thus providing a good sanity
check for the \ac{diagCCMC} approach. In addition, results at each level of theory are
expected to agree within statistical errors due to the size-consistency of all considered
approaches, as is observed.

\begin{table}[htb]
\caption{Correlation energy for different levels of theory using 1, 2 and 4
\ce{Be} replicas in a cc-pVDZ basis set. Note that for these systems CCSDTQ is
equivalent to \ac{FCI}. Molecular integrals were generated in FCIDUMP format
with the Q-Chem program package\cite{Shao2015-dy}. The canonical \acl{HF}
orbitals for a single-atom calculation were used, and no spin symmetry
breaking was observed, giving $E_{\mathrm{ref}} = \SI{-14.572341}{\hartree}$.}
\label{tab:ncombos-w-tl}
\begin{tabular}{l r S S S}
  \toprule
             & & \multicolumn{3}{c}{$n_{\mathrm{replicas}}$} \\
             & & 1 & 2 & 4 \\
  \cline{2-5}
  \multirow{2}{*}{SD}   & \ac{CC}\ac{MC} & -0.045032(2) & -0.09007(2) & -0.1799(1) \\
                         & \ac{diagCCMC}  & -0.04500(5)  & -0.09011(7) & -0.1801(3) \\
  \cline{2-5}
  \multirow{2}{*}{SDT}  & \ac{CC}\ac{MC} & -0.045067(2) & -0.09012(2) & -0.18034(7) \\
                         & \ac{diagCCMC}  & -0.04512(4)  & -0.0902(2)  & -0.1802(3) \\
  \cline{2-5}
  \multirow{2}{*}{SDTQ} & \ac{CC}\ac{MC} & -0.045070(2) & -0.09015(2) & -0.18044(9) \\
                         & \ac{diagCCMC}  & -0.04504(4)  & -0.0902(3)  & -0.1807(3) \\
  \cline{2-5}
  \ac{FCI}               &               & -0.0450721(7) & -0.090151(5) & -0.18036(6) \\
  \bottomrule
  \end{tabular}
\end{table}

In order to assess the computational performance of \ac{diagCCMC} we compare
two measures of efficiency:
\begin{itemize}
  \item $n_{\mathrm{attempts}}/\delta\tau$, that is, the number of stochastic samples performed
    per unit imaginary time. This metric is a measure of the \emph{minimum} CPU
    cost, provided that the length of propagation in imaginary time is roughly
    constant between approaches, or equivalently a roughly constant	inefficiency between
    the approaches.\citep{Vigor2016}
  \item $n_{\mathrm{states}}$, that is, the number of occupied excitation operators. This metric is
    a measure of the \emph{minimum} memory cost. For a deterministic calculation this would
    amount to the Hilbert space size for the selected truncation level.
\end{itemize}

%For the isolated and noninteracting replicas \ce{Be} systems, the results are
%plotted in Figures~\ref{fig:be-replica-nstates}
%and~\ref{fig:be-replica-nattempts} for $n_{\mathrm{states}}$ and
%$n_{\mathrm{attempts}}$, respectively.

\begin{figure}[htb]
\centering
\includegraphics[width=.7\textwidth]{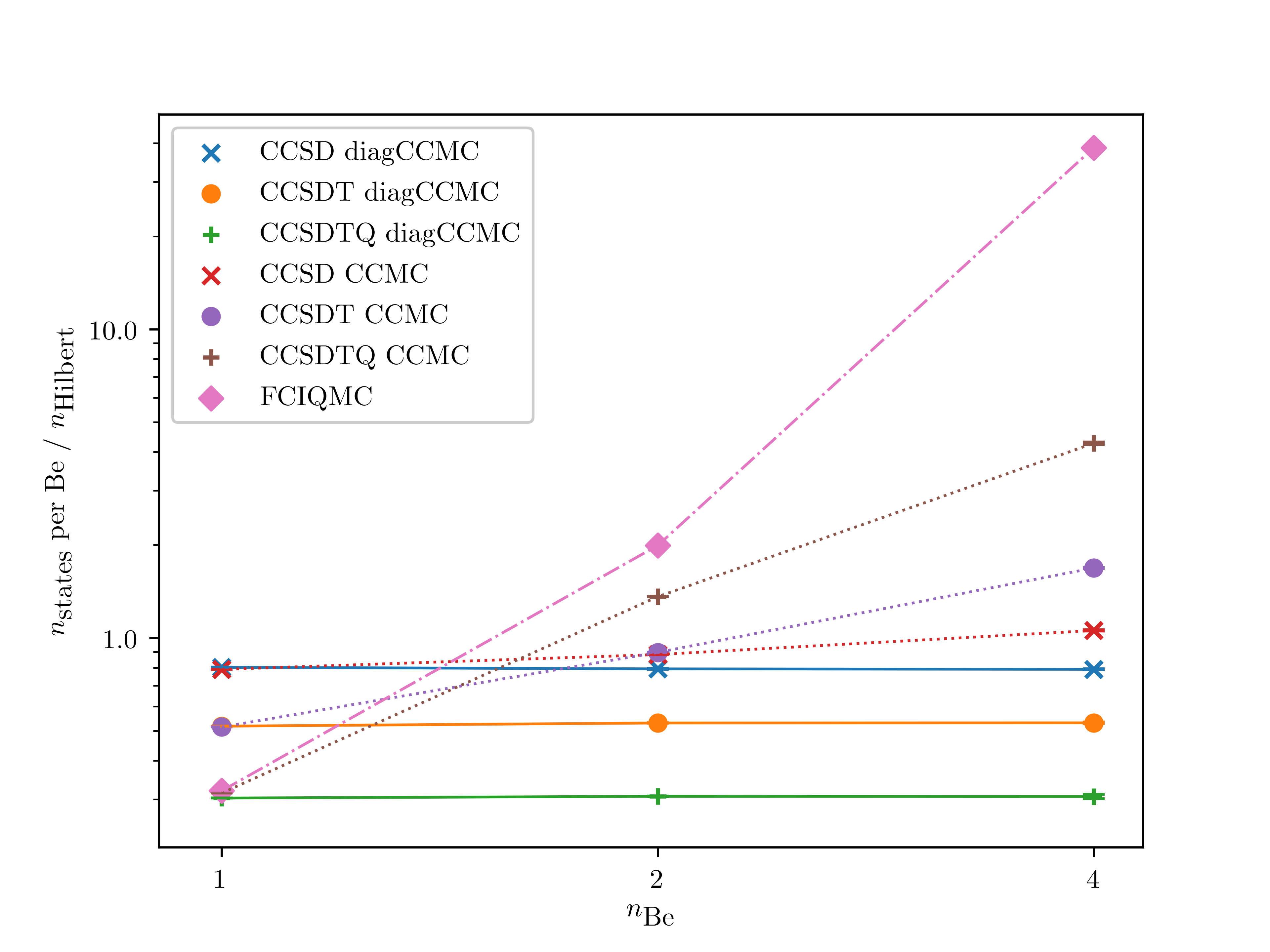}

\caption{
  The ratio of states per-replica and corresponding reduced Hilbert space size for 1, 2
  and 4 \ce{Be} replicas in a cc-pVDZ basis set at various levels of theory.
  The $n_{\mathrm{states}}$ metric is a measure of the memory cost of
  the calculation. For a single \ce{Be} atom the Hilbert space sizes are
  121, 529, and 1093 states for CCSD, CCSDT, and CCSDTQ, respectively, and
  the corresponding reduced Hilbert space multiplies these values by the
  number of \ce{Be} atoms.
  Note that for these systems CCSDTQ is equivalent to FCI.
  Solid, dotted and dash-dotted lines are used for \acs{diagCCMC},
  \ac{CC}\ac{MC} and \ac{FCIQMC} results, respectively.
  Molecular integrals were generated in FCIDUMP format with the Q-Chem program
  package\cite{Shao2015-dy}. The canonical \acl{HF} orbitals for a single-atom
  calculation were used, and no spin symmetry breaking was observed.
}
\label{fig:be-replica-nstates}
\end{figure}

The promise of stochastic methods is to greatly reduce the cost of high-level
correlated calculations by naturally exploiting the wavefunction sparsity.
Figure~\ref{fig:be-replica-nstates} reports the ratio of $n_{\mathrm{states}}$
per replica and the size of the Hilbert space for an isolated atom at the given
truncation level.
For an isolated \ce{Be} atom, the reduction in memory footprint is clearly evident: all methods compared require
significantly less than the full size of the Hilbert space (ratio $< 1$) to successfully
achieve convergence and recover the deterministic results. Unsurprisingly and
correctly, \ac{diagCCMC} requires the same amount of storage as its unlinked
counterparts. Notice also that the ratio decreases in going from CCSD to CCSDTQ
showing how stochastic methods single out the important portions of the Hilbert space.
For perfectly local systems, such as the noninteracting 2- and 4-atom replicas,
one also expects the number of states per replica to roughly stay constant.
This expectation stems from the linked diagram theorem\cite{Shavitt2009-mr} and
is met by the \ac{diagCCMC} approach where at each iteration only connected
diagrams are sampled.
The same is, quite emphatically, not true for either \ac{FCIQMC} or
\ac{CC}\ac{MC}: the number of states per replica approaches and surpasses the size of the
single-atom Hilbert space.

\begin{figure}[htb]
\centering
\includegraphics[width=.7\textwidth]{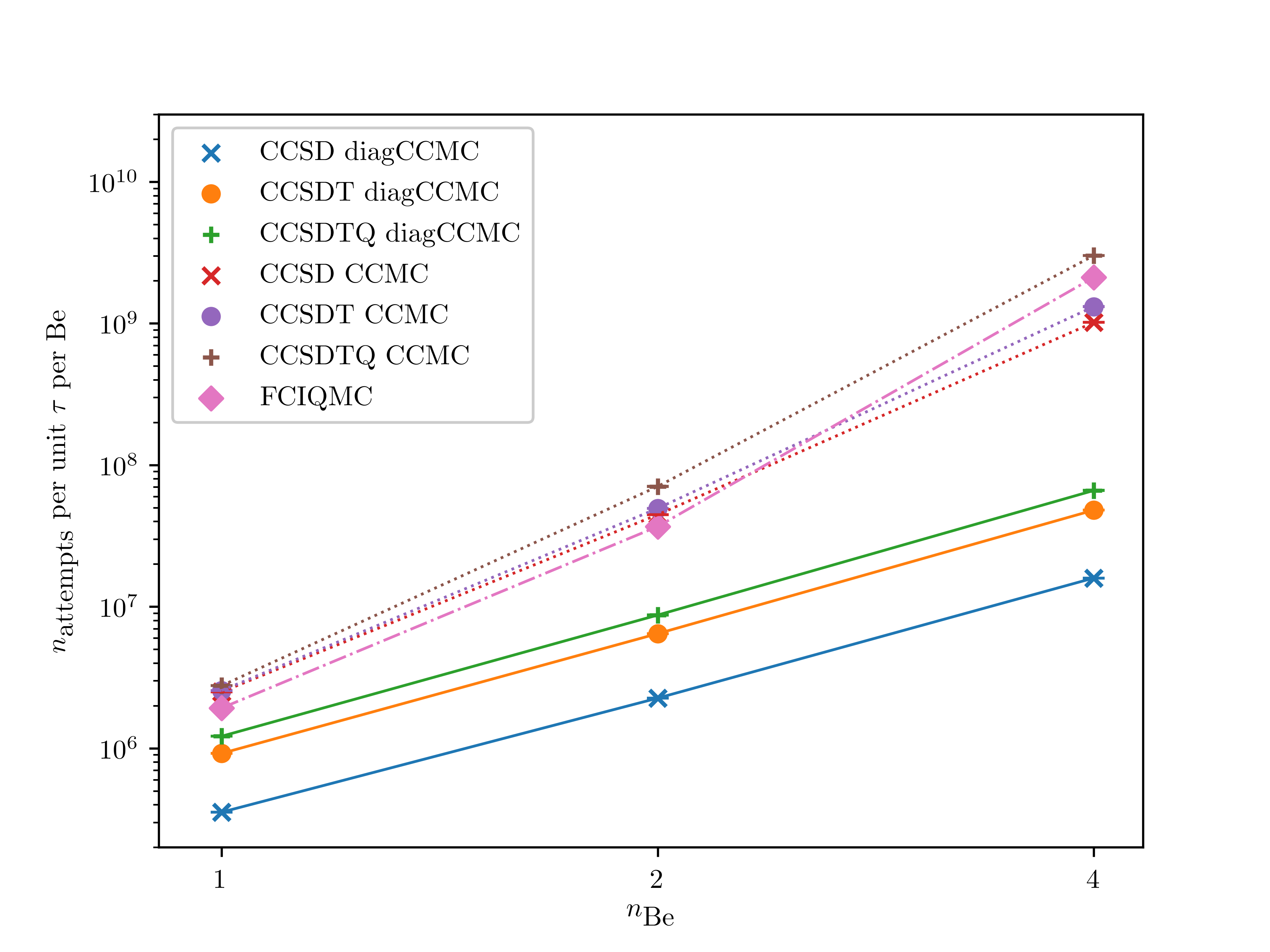}
\caption{
  Number of stochastic samples performed ($n_{\mathrm{attempts}}$) per unit imaginary time per
replica for 1, 2 and 4 \ce{Be} replicas in a cc-pVDZ basis set at various levels
of theory. Assuming that the length of propagation in imaginary time is roughly
constant between approaches this metric is a measure of the CPU cost of the calculation.
Note that for these systems CCSDTQ is equivalent to FCI.
Solid, dotted and dash-dotted lines are used for \acs{diagCCMC}, \ac{CC}\ac{MC} and \ac{FCIQMC} results, respectively.
Molecular integrals
were generated in FCIDUMP format with the Q-Chem program
package\cite{Shao2015-dy}. The canonical \acl{HF} orbitals for a single-atom
calculation were used, and no spin symmetry breaking was observed.
}
\label{fig:be-replica-nattempts}
\end{figure}

In Figure~\ref{fig:be-replica-nattempts} we can see that
\ac{diagCCMC} outperforms each of the corresponding \ac{CCMC} approaches also
when estimating the CPU cost of the calculations on the \ce{Be} systems here
considered. It is particularly striking to note the order of magnitude
difference between the diagrammatic and unlinked approaches at the \ac{CCSD}
level of theory even for this tiny system.

\begin{figure}[htb]
\centering
\includegraphics[width=.7\textwidth]{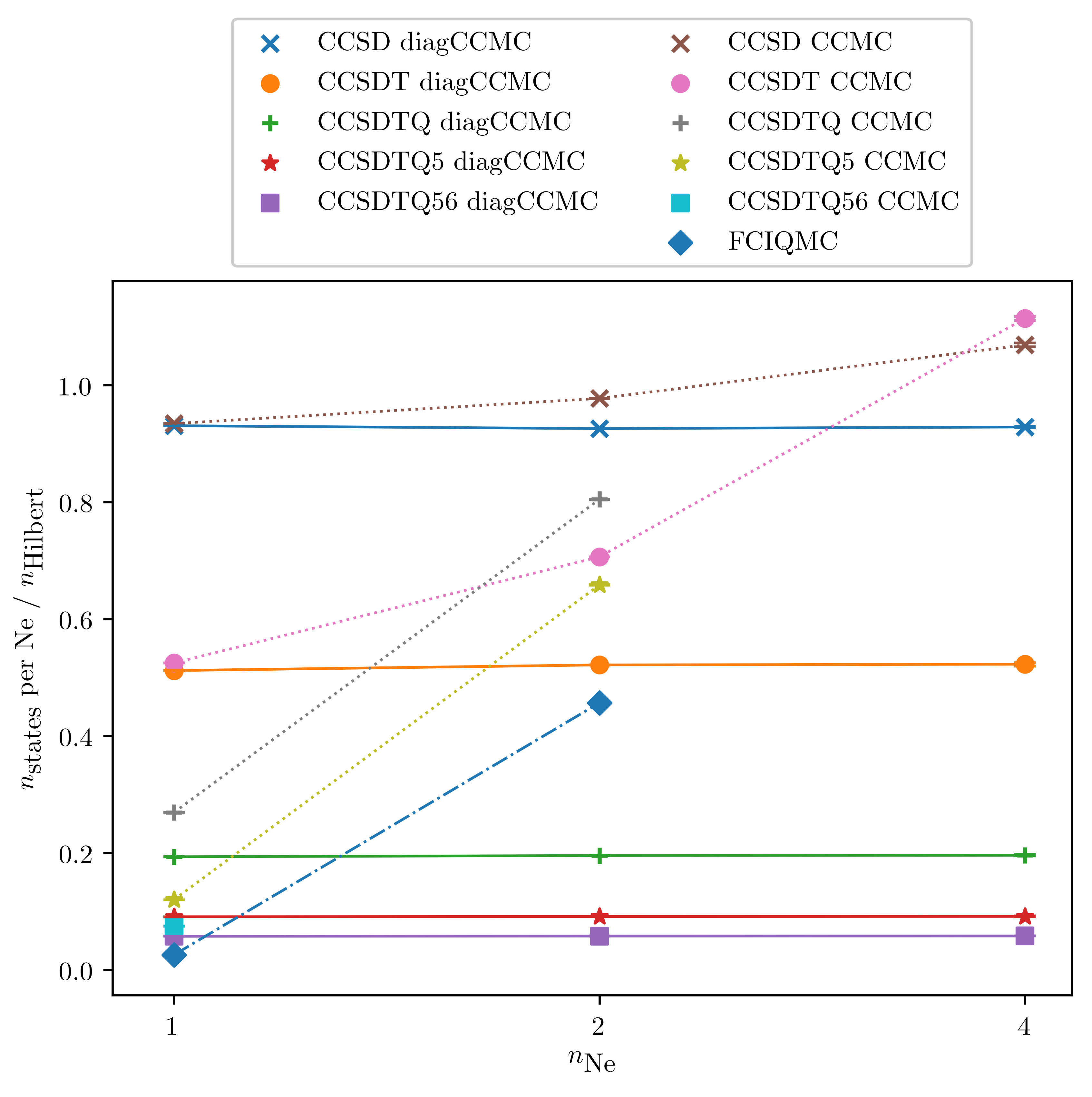}
\caption{
  The ratio of states per-replica and corresponding reduced Hilbert space size
  for 1, 2 and 4 \ce{Ne} replicas in a cc-pVDZ basis set at various levels of
  theory. The $n_{\mathrm{states}}$ metric is a measure of the memory cost of
  the calculation. For a single \ce{Ne} atom the Hilbert space sizes are 393,
  4647, 30861, 116129, 265790, 502099 for CCSD, CCSDT, CCSDTQ, CCSDTQ5,
  CCSDTQ56, and FCI, respectively, and the corresponding reduced Hilbert space
  multiplies these values by the number of \ce{Ne} atoms.
  Solid, dotted and dash-dotted lines are used for
  \acs{diagCCMC}, \ac{CC}\ac{MC} and \ac{FCIQMC} results, respectively.
  Molecular integrals were generated in FCIDUMP format with the Q-Chem program
  package\cite{Shao2015-dy}. The canonical restricted \acl{HF} orbitals for a single-atom calculation were used.
}
\label{fig:ne-replica-nstates}

\end{figure}

The same observation also holds true for higher orders of \ac{CC} theory, as can
clearly be seen from Figure~\ref{fig:ne-replica-nstates} where we plot the
$n_{\mathrm{states}}$ metric for an isolated \ce{Ne} atom and its
corresponding 2 and 4 noninteracting replicas system. Table~\ref{tab:neon}
reports the correlation energies per replica for a systems of noninteracting
\ce{Ne} atoms.
\ac{diagCCMC} affords calculations practically at constant memory cost per
replica in contrast with \ac{CC}\ac{MC} for which the increasing cost exceeded
available computational resources for the higher order excitations.

\begin{table}[htb]
\caption{
	Correlation energy for different levels of theory using 1, 2 and 4
	\ce{Ne} replicas in a cc-pVDZ basis set.
	Molecular integrals were generated in FCIDUMP format
	with the Psi4 program package\cite{Parrish2017-fd} and exact \ac{CC}
	results obtained using MRCC.\cite{MRCC} The canonical \acl{HF}
	orbitals for a single-atom calculation were used and no spin symmetry
	breaking was observed, giving $E_{\mathrm{ref}} = \SI{-128.488776}{\hartree}$.}
\label{tab:neon}
\begin{tabular}{l r S S S }
  \toprule
  		& & \multicolumn{3}{c}{$n_{\mathrm{replicas}}$} \\
  		& & 1 & 2 & 4 \\
  \cline{2-5}
  \multirow{2}{*}{SD}     & \ac{CC}\ac{MC} & -0.190865(3) & -0.38172(3) & -0.7633(2) \\
                         & \ac{diagCCMC}  & -0.19094(5)  & -0.3817(1) & -0.7641(5) \\

                         & \ac{CC} & -0.190861  & -0.38172$\text{3}^{\text{b}}$ & -0.76344$\text{6}^{\text{b}}$ \\
  \cline{2-5}
  \multirow{2}{*}{SDT}    & \ac{CC}\ac{MC} & -0.191951(4) & -0.38389(4) & -0.7676(3) \\
                         & \ac{diagCCMC}  & -0.19185(10) & -0.3839(2) & -0.7685(7) \\
                         & \ac{CC} & -0.191945  & -0.38389$\text{1}^{\text{b}}$ & -0.76778$\text{1}^{\text{b}}$ \\
  \cline{2-5}
  \multirow{2}{*}{SDTQ}   & \ac{CC}\ac{MC} & -0.192092(4) & -0.38418(6) & \hspace{1cm}\text{$\phantom{t}^{\text{a}}$} \\
                         & \ac{diagCCMC}  & -0.1924(1)   & -0.3843(6) & -0.7668(7) \\
                         & \ac{CC} & -0.192095  & -0.38419$\text{1}^{\text{b}}$ & -0.76838$\text{2}^{\text{b}}$ \\
  \cline{2-5}
  \multirow{2}{*}{SDTQ5}  & \ac{CC}\ac{MC} & -0.192103(4) & -0.38436(9) & \hspace{1cm}\text{$\phantom{t}^{\text{a}}$}  \\
                         & \ac{diagCCMC}  & -0.1924(2)   & -0.3840(5) & -0.7686(5) \\
                         & \ac{CC} & -0.192106  & -0.38421$\text{2}^{\text{b}}$ & -0.76842$\text{4}^{\text{b}}$ \\
  \cline{2-5}
  \multirow{2}{*}{SDTQ56} & \ac{CC}\ac{MC} & -0.192119(5) &  \hspace{1cm}\text{$\phantom{t}^{\text{a}}$} & \hspace{1cm}\text{$\phantom{t}^{\text{a}}$} \\
                         & \ac{diagCCMC}  & -0.1919(1)   & -0.3846(6) & -0.7691(5) \\
                         & \ac{CC} & -0.192106  & -0.38421$\text{1}^{\text{b}}$ & -0.76842$\text{2}^{\text{b}}$ \\
  \cline{2-5}
  \ac{FCI}                &               & -0.192106(5) & \hspace{1cm}\text{$\phantom{t}^{\text{a}}$} & \text{$\phantom{t}^{\text{a}}$} \\
  \bottomrule
  \end{tabular}

 \text{$\phantom{t}^{\text{a}}$} Value not computed due to computational constraints.\hfill

 \text{$\phantom{t}^{\text{b}}$} Value obtained as multiple of single atom result for comparison.\hfill
\end{table}

\begin{figure}[htb]
\centering
\includegraphics[width=.7\textwidth]{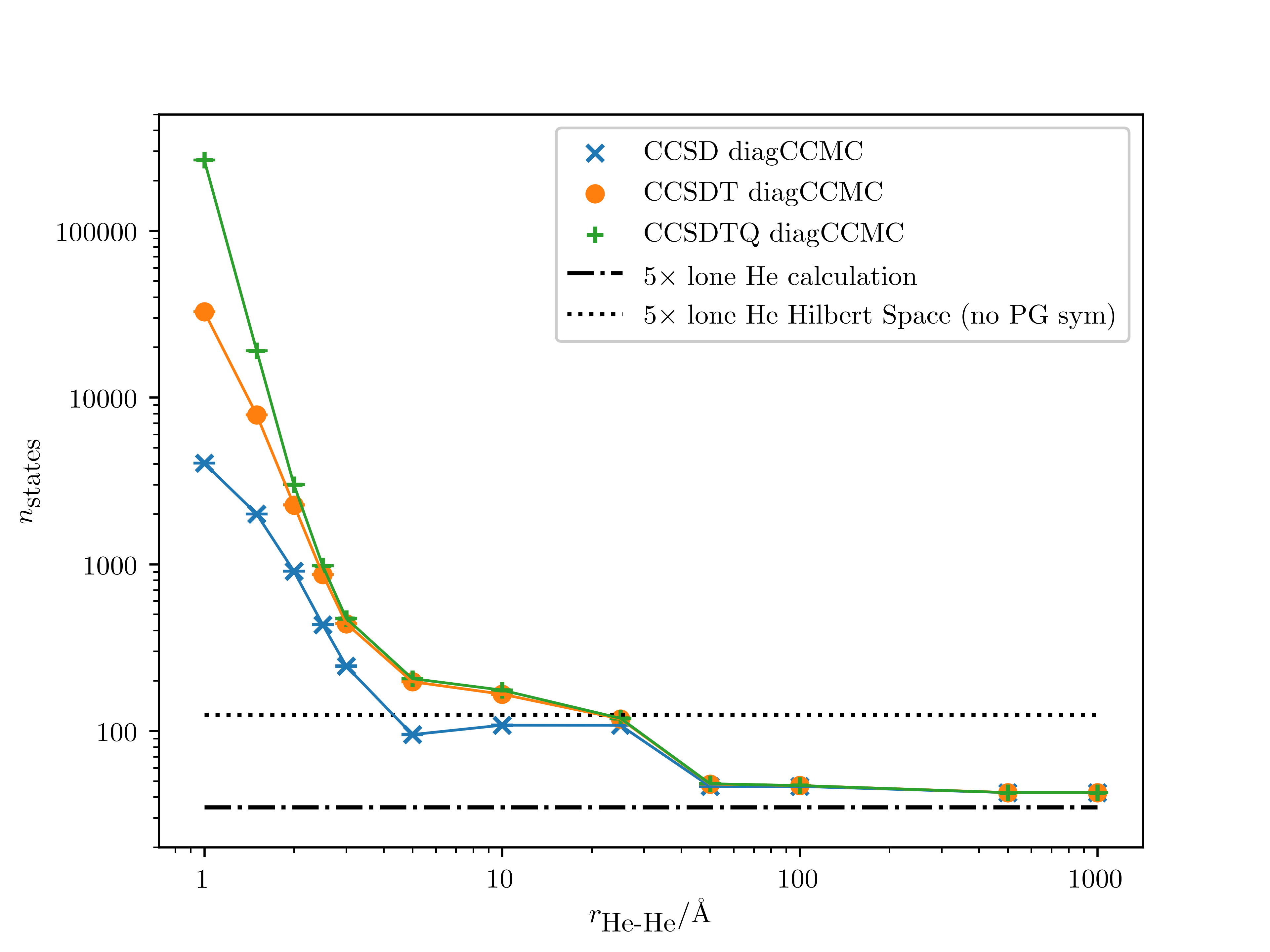}
\caption{The number of states ($n_{\mathrm{states}}$) for a line of 5 \ce{He}
atoms in a \ac{diagCCMC} calculation at the CCSD, CCSDT, and CCSDTQ levels of theory.
The $n_{\mathrm{states}}$ metric is a measure of the memory cost of the
calculation. Molecular integrals were generated in FCIDUMP format with the Psi4
program package\cite{Parrish2017-fd}. Localised orbitals were used: starting
from the restricted \acl{HF} canonical orbitals, the Foster--Boys\cite{Foster1960-md} and
the Pipek--Mezey\cite{Pipek1989-of} algorithms were used for the occupied and
virtual subspaces, respectively. For CCSD, we report 5$\times$ the Hilbert space size and
diagCCMC average memory cost for a single He atom using dash-dotted and
dotted horizontal lines, respectively.}
\label{fig:pent-he-dissoc-nstates}
\end{figure}

Finally, we studied the dissociation of a chain of 5 helium atoms as an example
of interacting system.
The diagrammatic algorithm shows favourable CPU and memory cost for
noninteracting systems, further suggesting that it might also straightforwardly
leverage localisation in the orbital space to achieve reduced cost for
calculations on interacting systems.
As a preliminary test for this conjecture,
Figure~\ref{fig:pent-he-dissoc-nstates} shows the memory cost for the
dissociation curve of an interacting chain of five helium atoms.
We localised the occupied and virtual orbital sets with the
Foster--Boys\cite{Foster1960-md} and the Pipek--Mezey\cite{Pipek1989-of}
criteria, respectively.
We compare the $n_{\mathrm{states}}$ metric with the memory cost at the
dissociation limit for a deterministic and a \ac{diagCCMC} CCSD calculation.
The former (dotted line) is the maximum memory cost for performing CCSD
calculations on the isolated atoms: below it, the cost is comparable to that for
a wavefunction with excitations localised to each \ce{He} atom. The onset of
such behaviour is evident from Figure~\ref{fig:pent-he-dissoc-nstates}, which
also shows the recovery of the noninteracting limit at large separations.

%% CONCLUSIONS

In conclusion, we have described a stochastic realisation of linked \ac{CC}
theory that fully exploits the connectedness of the similarity-transformed
Hamiltonian, as exemplified in the diagrammatic expansion of the \ac{CC}
equations.
Our stochastic diagrammatic implementation avoids the computational and memory
cost issues associated with deterministic and unlinked stochastic approaches, by
generating diagrams on-the-fly and accumulating the corresponding amplitudes.
Finally, we have shown how the stochastic and deterministic implementations can
be rationalised within the same framework. This bridges the existing gap between
the two strategies: by clearing possible misunderstandings on \emph{how} and
\emph{why} stochastic methods work and enabling future cross-fertilisation.

\begin{acknowledgement}

C.J.C.S. is grateful to the Sims Fund for a studentship and A.J.W.T. to the
Royal Society for a University Research Fellowship under Grant Nos. UF110161
and UF160398.
Both are grateful for support under ARCHER Leadership Project grant e507.
R.D.R. acknowledges partial support by the Research Council of Norway
through its Centres of Excellence scheme, project number 262695 and through its
Mobility Grant scheme, project number 261873.
R.D.R. is also grateful to the Norwegian Supercomputer Program through a grant
for computer time (Grant No. NN4654K).
T.D.C. was supported by grants CHE-1465149 and ACI-1450169 from the
U.S.~National Science Foundation.

\end{acknowledgement}

\begin{suppinfo}

Additional data related to this publication, including a copy of the
\ac{diagCCMC} code, raw and analysed data files and analysis scripts, is
available at the University of Cambridge data repository
\url{https://doi.org/10.17863/CAM.34952} and \url{https://doi.org/10.17863/CAM.36097}.

We used the \texttt{goldstone} \LaTeX~package, available on GitHub
\url{https://github.com/avcopan/styfiles}, to draw the \acl{CC} diagrams.
We used matplotlib for all the plots in the paper.\cite{Hunter2007-hv}

\end{suppinfo}

\bibliography{bibliography}

\end{document}